\documentclass[useAMS,usenatbib]{mn2e}
\usepackage{graphicx} 
\usepackage{times}
\usepackage{epsfig}
\usepackage{epstopdf}
\usepackage{amsfonts}
\usepackage{amsmath}
\usepackage{amsbsy}
\usepackage{bm}
\usepackage{url}
\usepackage{microtype}
\usepackage{rotating}
\usepackage{sidecap}
\usepackage{longtable}
\usepackage{lscape}
\usepackage{rotating}
\usepackage{color}
\usepackage{amssymb,xcolor}
\usepackage[normalem]{ulem}

\def\cm2{{\rm cm^{-2}}}
\def\ergcms{{\rm erg\,cm^{-2}\,s^{-1}}}

\newcommand{\swift}{\textsl{Swift}}
\newcommand{\inte}{\textsl{INTEGRAL}}

\title[V404 Cygni: an obscured AGN analogue]{The black hole binary V404 Cygni: a highly-accreting obscured AGN analogue}
\author[S. E. Motta et al.]{S. E. Motta$^{1}$, J.~J.~E. Kajava$^{2,3}$, C. S\'anchez-Fern\'andez$^{3}$, M. Giustini$^{4}$, E. Kuulkers$^{3,5}$  \\
$^{1}$University of Oxford, Department of Physics, Astrophysics, Denys Wilkinson Building, Keble Road, Oxford OX1 3RH, UK\\
$^{2}$Tuorla Observatory, University of Turku, V\"{a}is\"{a}l\"{a}ntie 20, FIN-21500 Piikki\"{o}, Finland \\
$^{3}$European Space Astronomy Centre (ESA/ESAC), Science Operations Department, 28691 Villanueva de la Ca\~nada, Madrid, Spain\\
$^{4}$SRON, Netherlands Institute for Space Research, Sorbonnelaan 2, 3584 CA Utrecht, The Netherlands\\
$^{5}$ESA/ESTEC, Keplerlaan 1, 2201 AZ Noordwijk, The Netherlands
\\
}
\begin{document}
\maketitle

\begin{abstract}

\noindent Typical black hole binaries in outburst show spectral states and transitions, characterized by a clear connection between the inflow onto the black hole and outflows from its vicinity. 
The transient stellar mass black hole binary V404 Cyg apparently does not fit in this picture. Its outbursts are characterized by intense flares and intermittent plateau and low-luminosity states, with a dynamical intensity range of several orders of magnitude on time-scales of  hours.
During the 2015 June-July X-ray outburst a joint \swift\ and \inte\ observing campaign captured V404 Cyg in one of these plateau states.
The simultaneous \swift/XRT and \inte/JEM-X/ISGRI spectrum is reminiscent of that of obscured/absorbed AGN. It can be modelled as a Comptonization spectrum, heavily absorbed by a partial covering, high-column density material ($N_\textrm{H} \approx 1-3 \times10^{24}\,\textrm{cm}^{-2}$), and a dominant reprocessed component, including a narrow Iron-K$\alpha$ line.
Such spectral distribution can be produced by a geometrically thick accretion flow able to launch a clumpy outflow, likely responsible for both the high intrinsic absorption and the intense reprocessed emission observed. 
Similarly to what happens in certain obscured AGN, the low-flux states might not be (solely) related to a decrease in the intrinsic luminosity, but could instead be caused by an almost complete obscuration of the inner accretion flow.

\end{abstract}

\begin{keywords}
Black hole - binaries: close - X-rays
\end{keywords} 



\section{Introduction}

Black hole (BH) X-ray binaries (BHBs) are typically transient systems that alternate between long periods of (X-ray) quiescence and relatively short outbursts. During the outbursts their luminosity increases by several orders of magnitude (from $\sim$10$^{32-34}$ erg/s in quiescence to $\sim$10$^{38-39}$ erg/s or more in outburst), due to an increase in the mass transfer rate to the BH. When active, most BHBs show an ``hysteresis'' behaviour that becomes apparent as cyclic loops in a so-called Hardness-Intensity diagram (HID; see e.g., \citealt{Homan2001}). 
These cyclic patterns have a clear and repeatable association with mechanical feedback in the form of different kind of outflows (relativistic jets and winds, see \citealt{Fender2009} and \citealt{Ponti2012}).

In a typical BHB different spectral-timing states can be identified with different areas of the q-shaped track visible in the HID. 
In the \textit{hard state} the X-ray energy spectrum is dominated by strong hard emission, peaking between $\sim$50-150 keV (e.g., \citealt{Sunyaev1979}, \citealt{Joinet2008}, \citealt{Motta2009}).
The likely radiative mechanism involved is Compton up-scattering of soft seed photons either produced in a cool geometrically thin accretion disk truncated at large radii, or by synchrotron-self-Compton emission from hot electrons located close to the central black hole (e.g., \citealt{Poutanen2014}).
In the \textit{soft state}, instead, the spectrum is dominated by thermal emission from a geometrically thin accretion disk that is thought to extend down or close to the innermost stable circular orbit (\citealt{Bardeen1972}). It is in this state that the peak X-ray luminosity is normally reached. 
In between these two states are the so-called \textit{intermediate} states, where the energy spectra typically show both the hard Comptonized component and the soft thermal emission from the accretion disk. 
In these states the most dramatic changes in the emission - reflecting changes in the accretion flow - can be revealed through the study of the fast-time variability (e.g., \citealt{Belloni2016}).

While most BHBs that emit below the Eddington limit fit into this picture, systems accreting at the most extreme rates do not.
A typical example is the BHB GRS 1915+105, which has been accreting close to Eddington during most of an on-going 23-years long outburst. 
Another example is the enigmatic high-mass X-ray binary V4641 Sgr (\citealt{Revnivtsev2002}), which in 1999 showed a giant outburst, associated to a super-Eddington accretion phase, followed by a lower accretion rate phase during which its X-ray spectrum resembled closely the spectrum of the well-known BHB Cyg X-1 in the hard state. While GRS 1915+105 displays relatively soft spectra when reaching extreme luminosities, V4641 Sgr did not, showing instead significant reflection and heavy and variable absorption, due to an extended optically thick envelope/outflow ejected by the source itself (\citealt{Revnivtsev2002}, \citealt{Morningstar2014}). 

When the accretion rate approaches or exceeds the Eddington accretion rate,  the radiative cooling time scale to radiate all the dissipated energy locally (a key requirement for thin disks) becomes longer than the accretion time scale. Therefore, radiation is trapped and advected inward with the accretion flow, and consequently both the radiative efficiency and the observed luminosity decrease. This configuration is known as \textit{slim disk} (\citealt{Begelman1979}, \citealt{Abramowicz1988}). The slim disk model has been successfully applied to stellar mass black holes, such as the obscured BHB candidate SS 433 (\citealt{Fabrika2004}), to ultraluminous X-ray sources (\citealt{Watarai2001}), and to super massive BHs (narrow-line Seyfert galaxies, e.g. \citealt{Mineshige2000}). 

High-accretion rate induced slim disks have been recently associated to high obscuration (high absorption) in a sample of weak emission-line AGN (\citealt{Luo2015}). In those sources, which are likely seen close to edge on, a geometrically thick accretion flow found close to the central supermassive BH is thought to screen the emission from the central part of the system, dramatically reducing the X-ray luminosity. Flared disks are also the most commonly used explanation for obscuration in X-ray binaries seen at high inclinations (see \citealt{White1982a} and, in particular, \citealt{Revnivtsev2002} for the case of V4641 Sgr, \citealt{Fabrika2004} for SS 433 and   \citealt{Corral-Santana2013} for Swift J1357.2--0933). 
In both the AGN and BH X-ray binary populations, a large fraction of faint (obscured), high-inclination sources seems to be missed by current X-ray surveys (e.g., \citealt{Ballantyne2006}, \citealt{Severgnini2011} and \citealt{Corral-Santana2013}).

Even considering the entire population of accreting sources as a whole - encompassing stellar mass objects (compact and not), Ultra-luminous X-ray sources (ULXs \citealt{Feng2011}) and active galactic nuclei (AGN)  - only a small fraction of the known systems seems to be accreting close to Eddington rates, one of them being V404 Cyg (\citealt{Zycki1999}). 
V404 Cyg is an intermediate to high-inclination (\citealt{Sanwal1996}), intrinsically luminous, likely often super-Eddington during outbursts (\citealt{Zycki1999}) confirmed BHB (\citealt{Casares1992}): studying this system opens the opportunity to probe a 
regime where high accretion rates, heavy and non-homogeneous absorption and reflection are interlaced and all play a key role in the emission from the source. Hence, understanding the physics of V404 Cyg's emission could shed light on the accretion related processes occurring not only in stellar mass BHs, but also in ULX sources  and, most importantly, in AGN.

\section{V404 Cyg, a.k.a. GS 2023+338}

V404 Cyg was first identified as a optical nova in 1938 and later  associated to the X-ray transient GS~2023+338, discovered by \textit{Ginga} at the beginning of its X-ray outburst in 1989 (\citealt{Makino1989}). The 1989 outburst displayed extreme variability, reaching several flux levels above that of the Crab. During this outburst, V404 Cyg became temporarily one of the brightest sources ever observed in X-rays. 
\citet{Casares1992} determined the orbital period of the system ($\sim$6.5 days) and \cite{Miller-Jones2009} the distance to the source through radio parallax (d = 2.39 $\pm$ 0.14 kpc). 
\cite{Casares1992} also obtained the first determination of the system's mass function ($f$(M) = 6.26 $\pm$ 0.31 M$\odot$), confirming the black hole nature of the compact object in V404 Cyg and allowing to classify it as a low-mass X-ray binary. \citet{Shahbaz1994} later determined a BH mass of about 12M$_{\odot}$. More recently, near-infrared spectroscopy allowed a more precise determination of the compact object mass, $M_\textrm{BH} = 9.0^{+0.2}_{-0.6}\,\textrm{M}_{\odot}$ \citep{Khargharia2010}.

On 2015 June 15 18:32 UT (MJD 57188.772), the \swift/BAT triggered on a bright hard X-ray flare from a source that was soon recognized to be the black hole low mass X-ray binary V404 Cyg  back in outburst after 26 years of quiescence (\citealt{Barthelmy2015}, \citealt{Kuulkers2015}).
V404 Cyg reached the outburst peak on June 26 and then began a rapid fading towards X-ray quiescence, that was reached between 2015 August 5 and August 21 (\citealt{Sivakoff2015}). All along this outburst the source displayed highly variable multi-wavelength activity (\citealt{Rodriguez2015}), that was monitored by the astronomical community through one of the most extensive observing campaigns ever performed on an X-ray binary outburst (see \citealt{Sivakoff2015}, and references therein).

Already during the 1989 outburst (e.g. \citealt{Oosterbroek1996}, \citealt{Zycki1999}), V404 Cyg seemed to break the typical BHB pattern. Since we now know the distance of V404 Cyg with high precision, we can say that in 1989 it showed luminosities exceeding the Eddington limit, but without showing a canonical disk-dominated state (however, see \citealt{Zycki1999}, who report on a short-lived disk-dominated state).
Furthermore, the outburst was characterized by extreme variability, partly due to mere accretion events (somewhat similar to those see in GRS 1915+105, see \citealt{Belloni1990}), but also ascribed to a heavy and strongly variable photo-electric local absorption (\citealt{Tanaka1995}, \citealt{Oosterbroek1996}, \citealt{Zycki1999}).

\section{Data reduction and Analysis}

After the initial \swift/BAT trigger (on 2015 June 15 18:32 UT, MJD~57188.772), \inte\ observed V404 Cyg almost continuously during its entire outburst, providing the best hard-X-ray coverage ever obtained for this source \citep{Kuulkers2015b, Kuulkers2015c}.
\swift\ provided several short observations (often more than one per day) from the start of the outburst all the way down to quiescence.
Analysis of the \inte/ISGRI spectra - where absorption has little effect - showed that the source was sometimes seen in a plateau state, where the spectra could be described solely by a pure reflection spectrum from neutral material \citep{Sanchez-Fernandez2016}. 
One \swift\ pointing (OBSID: 00031403048) happened to take place exactly during one of these states and, differently from what has been seen during the rest of the outburst (see, e.g. \citealt{Natalucci2015}), both the flux and the spectral shape of V404 Cyg were remarkably stable, allowing us to obtain an high-quality average broad band X-ray spectrum of the source. 

\subsection{\textit{INTEGRAL}}

\inte\ data were processed using the  \textit{Off-line Scientific Analysis} software (OSA, \citealt{Courvoisier2003}), v10.2, using the latest calibration files at the time of the analysis. We selected only those \inte\ data which were strictly simultaneous to the two \swift\ snapshots described below, by using the appropriate  good time interval (GTI) files. 
IBIS/ISGRI and JEM-X data were processed from the COR step to the SPE step, using standard reduction procedures. The IBIS/ISGRI spectra were extracted using the OSA default energy binning,  which samples the energy range 20--500\,keV using 13 channels with logarithmic variable energy bins. We fit the ISGRI spectrum between 20 and 250\,keV (above 250\,keV the emission is background dominated). The JEM-X spectra were extracted using 23 user-defined energy bins, adjusted to allow a better sampling of the energy region around the Iron-K$\alpha$ line.
We modelled the final JEM-X spectrum between 5 and 25\,keV,  using 16 energy channels, given the uncertainties in calibration outside this band.
The IBIS/ISGRI and JEM-X spectra extracted to coincide with the two \swift\ snapshots were subsequently combined in a single spectrum per instrument. The net (dead-time corrected)  exposure times  for the combined spectra were 678\,sec for IBIS/ISGRI and 866\,sec for JEM-X. To account for calibration uncertainties, 5 per cent systematic errors was added to both spectra.

\subsection{\textit{Swift}}\label{sec:data_red}

Observation 00031403048 was taken in Window Timing (WT) mode on 2015-06-21 at 03:55:18 UTC and had a total exposure of 994s, split in 2 snapshots. We extracted events in an circular region centred at the source position with fixed outer radius (30 pixels, \textit{source region} from now on). To produce the energy spectrum we considered only grade 0 events and ignored data below 0.6\,keV in order to minimize the effects of high absorption and possible effects of residual pile-up. However, the average count rate of this observation in the source region was just above 10 counts per second in the 0.6--10\,keV band, therefore pile-up is unlikely (see http://www.swift.ac.uk/analysis/index.php).
Since the spectra extracted from the first and second snapshots of observation 00031403048 did not show significant differences, we produced one single spectrum from the entire observation to improve the signal to noise ratio. We fitted the combined XRT spectrum between 0.6 and 10\,keV.

\subsubsection{Treatment of the dust scattering X-ray halo}

As reported by \cite{Vasilopoulos2015}, \cite{Heinz2016} and \cite{Beardmore2016}, in some of the \swift\ images an X-ray halo caused by interstellar dust scattering is seen around the source. This halo emission may strongly contaminate the background region used for spectral extraction. For this reason, we used in our fits an alternative background file extracted from a routine WT mode calibration observation of RXJ1856.4-3754 in March 2015 (exposure 17.8\,ks).
The background was extracted from an annular region centred at (RA, Dec = 284.17, -37.91\,degrees) with inner and outer radii of 80 and 120 pixels.  

The X-ray halo may also contaminate the spectrum of V404 Cyg, producing an excess in soft X-rays.  To evaluate this possible contamination, we also extracted the spectrum of the dust-scattering halo using an annular region
centred at the source position with inner and outer radius 30 and 60 pixels away from the source position, respectively (\textit{halo region} from now on). We fitted the halo spectrum over the energy range 0.8 -- 4\,keV. This energy range was selected to avoid the possible distortion in the spectrum at energies below $\sim$1\,keV\footnote{Below $\sim$1\,keV the fact that the scattering cross section significantly drops off towards low energies (e.g. \citealt{Smith1998c}) can significantly distort the halo spectrum. } and to avoid the energy band where the X-ray background might start to dominate over the halo spectrum. 
Both the source and halo spectral channels were grouped in order to have a minimum number of 20 counts per bin. A 3 per cent systematic error was added to both spectra.

The X-ray spectrum of the dust scattering halo can be well-described by a soft power law (photon index $\Gamma \sim 3$),\footnote{The slope of the power law component describing the dust scattering halo spectrum depends on the spectral shape of the emission that caused the particular halo ring that might be contaminating the central source at a given time; Beardmore et al., private communication} affected by interstellar absorption (the same that does affect the source).
The halo is variable on a time-scale significantly longer than the characteristic source variability time-scales, and can be considered constant during the XRT observation analysed here. 

\begin{figure*}
\centering
\includegraphics[width=1\textwidth]{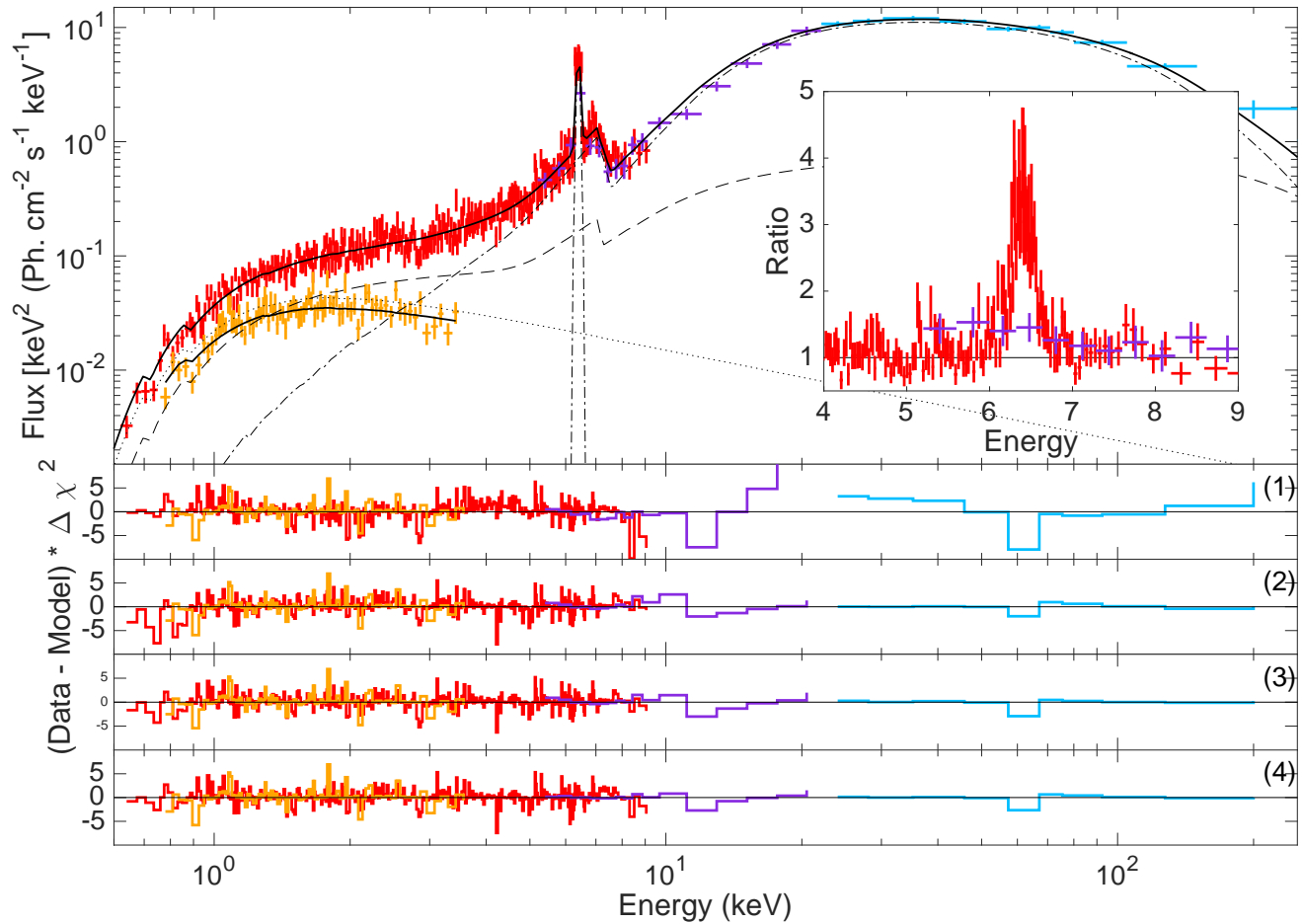}
\caption{\swift/XRT + \inte/JEM-X + \inte/ISGRI spectrum of V404 Cyg. The spectra were extracted in the good-time interval obtained from the \swift\ observation 00031403048, taken on 2015-06-21 at 03:55:18 UTC with an exposure of 994s. 
\textit{Upper panel:} \swift/XRT (red), \inte/JEM-X (purple), \inte/ISGRI (clear blue) source spectra and \swift/XRT halo spectrum (orange). The solid line marks best fit (model 3, see text) to the source and halo spectra. The dot-dashed line marks the reflection spectrum (including the Gaussian line), the dashed line marks the Comptonization spectrum and the dotted line marks the halo spectrum (in the source region, see text for details). \textit{Lower panels:} residuals to the model described in the text. From top to bottom: Model 1 (partially absorbed Comptonization spectrum), Model 2 (partially absorbed pure reflection spectrum), Model 3 (partially absorbed Comptonization with reflection spectrum), Model 4 (partially absorbed Comptonization with reflection spectrum, with variable local absorber) A colour version of this figure is available on-line. The inset plot shows the ratio of model to data once the Gaussian line has been removed from the model. Only the Iron-K$\alpha$ line region is shown.}
\label{fig:spettro}
\end{figure*}

In order to properly disentangle the halo emission and the source emission, we simultaneously fit the source spectrum and the dust scattering halo spectrum (see Fig. \ref{fig:spettro}) using an absorbed power law to describe the X-ray halo emission (absorption tied to the interstellar value). Since the contribution of the halo emission to the source region can be in principle different from the contribution of the halo emission to the background, we left the normalization of the power law component describing the halo free to vary, while the photon index was set by the halo spectrum alone. 

The models described below will therefore have the form \textsc{cons*tbnew1*[source model] +tbnew2*powerlaw} in \textsc{XSPEC}, where \textsc{cons} is a calibration constant (fixed at 1 for the \swift/XRT spectra and left free for the ISGRI and JEM-X spectra) and where the second term of the expression is aimed at fitting only the dust scattering halo emission. The absorption applied to the halo (\textsc{tbnew2}) is tied to the absorption  applied to the source spectrum (\textsc{tbnew1}), which is fixed at 8.3 $\times$ 10$^{21}$ cm$^{−2}$ (\citealt{Valencic2015}). 
Different \textsc{source  model} options were tested on our data-set. These are
described in detail in Sec. \ref{sec:Comp} to \ref{sec:Comp_refl} (Model 1) and in \ref{sec:model2} (Model 2). All the free parameters derived from these fits are summarized in Tab. \ref{tab:parameters} for Model 1 and in Tab. \ref{tab:parameters2} for Model 2. We used $\chi^2$ statistics in the model selection and for parameter error determination. In the following, we quote statistical errors at the 1$\sigma$ confidence level ($\Delta \chi^2=1$ for one parameter of interest).

\begin{table*}
\centering
\caption{Best fitting parameters for the three cases of Model 1 described in the text. {\bf Case} 1: partially absorbed Comptonization; {\bf Case 2:} partially absorbed pure reflection spectrum; {\bf Case 3:} partially absorbed Comptonization spectrum with reflection; {\bf Case 4:} partially absorbed Comptonization with reflection and variable local absorber . Parameters are: intrinsic column density associated to the direct (Compton) spectrum and relative covering fraction (pcf${NH1}$ and pcf$_{NH1}$),  intrinsic column density associated to the reflection spectrum and relative covering fraction (pcf${NH1}$ and pcf$_{NH1}$), electron temperature, Compton parameter (see text), relative reflection factor, ionization parameter (see text), \textsc{compps} normalization, Gaussian line energy, Gaussian line equivalent width, power law photon index (for the dust scattering halo), power law normalization (for the dust scattering halo), JEM-X calibration constant, ISGRI calibration constant. All the quoted errors are 1$\sigma$ level ones. Notice that all models include an  additional absorber to account for the interstellar absorption ($N_\textrm{H} = 8.3 \times 10^{21}\,\textrm{cm}^{-2}$). The Fe K-$\alpha$ line flux has been expressed in term of erg cm$^{-2}$s$^{-1}$ instead of equivalent width in order to allow direct comparison with the lines flux from Model 2 (see \ref{sec:model2}, for which measuring the equivalent width is problematic.)}\label{tab:parameters}
\begin{tabular}{c c c c c}

\hline
Model 1 parameter		& Case 1					&	Case 2				  & Case 3       		& Case 4       				 \\
\hline
\hline		 \\

Intrinsic $N_\textrm{H1}$ [cm$^{-2}$]& (140 $\pm 8$)$\times$10$^{22}$ 			&	-										 & ($139_ \pm 13 $)$\times$ 10$^{22}$ & 	($29_{-6}^{+7}$)$\times$ 10$^{22}$					 \\
pcf$_{NH1}$           				 & 0.960 $\pm 0.005$               			& 	-	   							 		 & $0.85 \pm 0.03$                    & $0.72 _{-0.06}^{+0.05}$        \\
Intrinsic $N_\textrm{H2}$ [cm$^{-2}$]& -							 			&	($157 \pm 14$)$\times$10$^{22}$  		 & $= N_\textrm{H1}$ 	   			  & ($136_{-16}^{+18}$)$\times$ 10$^{22}$ \\
pcf$_{NH2}$           				 & -	& $0.83 \pm 0.03$	   		  		& = pcf$_{NH1}$        						& 1 (fixed)        										\\
$kT_\textrm{e}$ [keV]    			 & $22 \pm 1$              	   	  			& $87_{-16}^{+24}$ 			     			& $65_{-10}^{+15}$                 & $57.61_{-8}^{+11}$   		\\
$\textit{y} $       				 & $1.16_{-0.08}^{+0.1}$     	  			& $0.92_{-0.1}^{+0.09}$ 		     		& $1.00_{-0.09}^{+0.08}$           & $0.97_{-0.07}^{+0.07}$       \\
\textit{Refl}	        			 & 0 (fixed)                 	  			& -1 (fixed)				  		 		& $19_{-5}^{+7}$                   & $31_{-8}^{+1}$    	\\
$\xi$ [erg cm s$^{-1}$]  	     	 & 0  (fixed)	    		      			& $6_{-4}^{+9}$ 			 				& $9_{-6}^{+7}$              	   & 0  (fixed)	 		\\
K$_{CompPS}$     					 & ($1.9_{-0.4}^{+0.5}$)$\times$10$^{6}$   	&($1.1_{-0.3}^{+0.4}$)$\times$10$^{7}$ 		& (4 $\pm 1$)$\times 10^{5}$       & (4 $\pm 1$)$\times 10^{5}$		\\
Line Energy [keV]   				 & $6.40 \pm 0.05$         			    	& $6.40 \pm 0.01$			         		& $6.40 \pm  0.05$   			   & $6.40 \pm  0.01$ 	    \\

Fe K-$\alpha$ Line flux [erg cm$^{-2}$s$^{-1}$] & (7.8$\pm$0.8)$\times$10$^{-10}$ & (6.3$\pm$0.6)$\times$10$^{-10}$     	& (6.0$\pm$0.6)$\times$10$^{-10}$  &  (6.0$\pm$0.6)$\times$10$^{-10}$	    \\

$\Gamma_\textrm{halo}$				 & $2.82 \pm 0.06$                   		& $2.63 \pm 0.08$  			  				& $2.79 \pm 0.06$		  		 	& $2.81 \pm 0.06$		\\
K$_\textrm{halo}$  					 & ($3.1 \pm 0.1$) $\times$10$^{-2}$  &($1.6 \pm 0.5$) $\times$10$^{-1}$	  			& (9 $\pm 1$)$\times$ 10$^{-2}$		& (5 $\pm 1$)$\times$ 10$^{-2}$		\\
cross-normalization constant (JEM-X) & $1.42  \pm 0.07$                   	& $1.3 \pm 0.2$ 			  				& 1.27 $\pm 0.07$  			         	& 1.7 $\pm 0.2$	\\
cross-normalization constant (ISGRI) & $2.5_{-0.4}^{+0.3}$            		& $1.5 \pm 0.5$		  						& $1.5_{-0.1}^{+0.2}$ 	  	         	& $4_{-2}^{+3}$ 	\\
\hline
$\chi^2$							 & 620.12    							&	499.77				  					& 474.28     				         	& 482.98	\\
Degrees of freedom					 & 461    								&	460				  						& 459				       	         	& 459	\\
Null hypothesis probability 		 & 9.82$\times$10$^{-8}$    			&	9.73$\times$10$^{-2}$		  			& 3.02$\times$10$^{-1}$			 		& 2.12$\times$10$^{-1}$	\\

\hline
\end{tabular}
\end{table*}

\begin{table}
\centering
\caption{Fluxes obtained in the 0.6-200 keV energy band from the best fit from Model 1, case 3, see Sec. \ref{sec:Comp_refl}).  We report both the observed fluxes (corrected only for the interstellar equivalent column density) and the intrinsic ones, corrected for the absorption local to the source. For reference, we note that for a stellar mass black hole of 9 M$\odot$ and at 2.39 kpc, as appropriate for V404 Cyg, the flux corresponding to the Eddington luminosity equals 1.6$\times 10^{-6}\, \ergcms$. }\label{tab:flux_comp_refl}
\begin{tabular}{c c }		
\hline	  
Component 					& 	Flux [erg/cm$^{2}$/s]		\\
\hline
\hline

Source  (observed)						& 3.77 $\times 10^{-8}$ 	\\
Source  (intrinsic)  		 			& 4.46 $\times 10^{-8}$ 	\\
Compton continuum (observed) 			& 3.46 $\times 10^{-9}$ 	\\
Compton continuum  (intrinsic)			& 6.51 $\times 10^{-9}$ 	\\
Reflection (absorbed) 					& 3.42 $\times 10^{-8}$ 	\\
Reflection  (intrinsic) 		 		& 3.80 $\times 10^{-8}$ 	\\
Halo$^*$ (in the source region)	 		& 2.76 $\times 10^{-10}$ 	\\

\hline
\end{tabular}
\end{table}


\section{Spectral modelling - Model 1: partially covered reflected Compton spectrum}\label{sec:model1}

\subsection{Case 1: Partially absorbed Comptonization}\label{sec:Comp}

We initially fitted our data using the Comptonization model \textsc{compps} in \textsc{XSPEC} (see \citealt{Poutanen1996}) modified by the interstellar absorption (\textsc{tbnew} in \textsc{XSPEC}, \citealt{Wilms2000}) with a fixed column density of $N_\textrm{H} = 8.3 \times 10^{21}\,\textrm{cm}^{-2}$.
We left the Thomson optical depth $\tau$ electron temperature $kT_\textrm{e}$ and the normalization in \textsc{compps} free to vary, while we fixed the seed photon temperature to $kT_\textrm{bb} = -0.1$ keV (i.e. the seed photons are produced by a multi-color disk-blackbody with an inner disk temperature of 0.1\, keV) and the inclination to 67$^{\circ}$ (see \citealt{Khargharia2010}). We also fixed to zero the ionization parameter $\xi$ since it only affects the reflection component, which is switched off in this case.  
We left all the remaining \textsc{compps} parameters fixed to their default values.
Large residuals indicated that a more complex spectral model was required. 
Therefore, we added a neutral absorber partially covering the source (\textsc{tbnew\_pcf}).
The addition of a narrow Gaussian line to describe the Iron-K$\alpha$ line (see \citealt{King2015}) was required to account for some residuals around 6.4 keV. The line is unresolved in both the \swift/XRT and \inte/JEM-X spectra, therefore we fixed its width to 0.2 keV.
This model did not provide satisfactory fits, having $\chi^2$ = 620.12 with 461 d.o.f.\ and null hypothesis probability = 9.82$\times$10$^{-8}$. Large residuals especially above 10 keV indicate that the spectrum likely shows a significant reflection component.  

\subsection{Case 2: Partially absorbed pure reflection}\label{sec:Pure_refl}

Since the emission from the dust scattering halo was at times significant in the sky region around the source, sometimes outshining the source itself (see \citealt{Vasilopoulos2015}, and \citealt{Heinz2016}, \citealt{Beardmore2016}), we also fitted our broad-band spectrum with a pure-reflection spectrum (\textsc{compps} with reflection scaling fraction parameter - defined as $\mathrm{Refl}$ = $\Omega/2\pi$ - frozen at -1, combined with a Gaussian line), superimposed to a steep power law used to model the X-ray halo spectrum, as described in Sec. \ref{sec:data_red}.
Again, we fitted simultaneously the dust scattering halo and the source spectrum to constrain the slope of the power law associated with the halo, while leaving its normalization free to vary. We did not find any evidence of a soft excess that required the addition of a soft component (e.g. a disk-blackbody) to our model. The best fit parameters are reported in Tab. \ref{tab:parameters} (Case 2).

It is worth noticing that this best fit corresponds to an unlikely halo flux in the source region equal to 3$\times$10$^{-10}$ erg/cm$^2$/s, i.e. more than a factor of 2 higher than the flux in the halo region (1.4$\times$10$^{-10}$ erg/cm$^2$/s).
Furthermore, when fitting our spectra with this model, the power-law photon index (associated with the dust halo) is significantly smaller than the photon index derived in case 1, possibly as a consequence of the presence of negative residuals at very low energies (below 1\, keV) in the source spectrum. This causes the development of more residuals in the dust halo spectrum around 4\, keV, which suggest that the photon index is likely forced to assume lower values by the source spectrum. The photon index derived using this model is probably too low to properly describe the halo spectrum, which  further points to a non accurate spectral modelling.

\subsection{Case 3: Partially absorbed Comptonization and significant reflection}\label{sec:Comp_refl}

Finally, we added a reflected component to the model described in Sec. \ref{sec:Comp}, by allowing the reflection scaling factor parameter in \textsc{compps} to vary freely. As in the previous models, we fitted simultaneously the source and the halo spectra. Also using this model there is no signature of a soft excess requiring the use of an additional soft component, while if we let the seed photon temperature vary, it remains consistent with 0.1 keV. We also note that the normalization of the \textsc{compps} component corresponds to an apparent inner disk radius of about 10$R_g$. 
According to the best fit, the reflected component of the spectrum provides $\sim$90\% of the total unabsorbed source emission. 
The best fit gives $\chi^2$ = 474.23, 459 d.o.f.\ and null hypothesis probability = 0.302. The best fit parameters are reported in Tab. \ref{tab:parameters} (Case 3). We list in Table \ref{tab:flux_comp_refl} the fluxes measured in the 0.6--200 keV energy range.
From our best fit we find that the halo contributes to 0.7\% of the total flux (source + halo) in the source region.

This model is strongly statistically favoured with respect to the model described in Case 1 (see Sec. \ref{sec:Comp}): an \textit{F-TEST} returns an F-statistic value = 70.60 and null hypothesis probability 1.85$\times$10$^{-27}$. On the other hand, while the model used in Case 2 (see Sec. \ref{sec:Pure_refl}) gives an acceptable fit to the data with $\chi^2$ = 499.77, 459 d.o.f.\ and null hypothesis probability = 0.0973, it is still statistically disfavoured with respect to the model described in this section. An \textit{F-test} returns F statistic value = 24.71 and probability 9.39$\times$10$^{-7}$ in favour of the model including both a direct Comptonization spectrum and a reflected component. 

\subsection{Case 4: Partially absorbed Comptonization and significant reflection, with variable local absorber }\label{sec:Comp_refl_2pcf}

Since it is entirely possible that the illuminating spectrum and reflected emission are produced in different regions of the accreting system, it is also possible that the two components are affected in different ways by the heavy absorption that partially covers the source, which has been treated so far as an average quantity. Therefore, we attempted to separate the average local column density into two components, applied to the illuminating spectrum (N$_{H1}$) and to the reflected one (N$_{H2}$), respectively. In order to avoid spectral degeneracy, we fixed the partial covering fraction associated to the reflected spectrum, pcf$_{NH2}$, to 1 (i.e. uniform covering) while leaving the partial covering fraction associated to the direct spectrum, pcf$_{NH1}$, free to vary. Since the ionization parameter $\xi$  was \textit{pegged} at zero during the spectral fitting, we fixed it to zero during the spectral fit.

The best fit gives $\chi^2$ = 482.98, 459 d.o.f.\  and null hypothesis probability 0.219. The best fit parameters are reported in Tab. \ref{tab:parameters} (Case 4). While this model gives an acceptable fit to the data, it is  statistically disfavoured with respect to the model described in Sec. \ref{sec:Comp_refl}, which remains our best fit so far. We note, however, that the constant taking into account the instrumental cross-calibrations are not as well-constrained as in case 2 and 3, which could indicate the presence of mild spectral degeneracy.

An obvious extension to this model would be to leave the partial covering fraction parameter of the absorber associated to the reflected spectrum (pcf$_{NH2}$) free to vary. However, this results into pcf$_{NH2}$ being \textit{pegged} at 1, while the remaining parameters are consistent with those reported in Tab. \ref{tab:parameters}, case 4.


\subsection{Iron-K$\alpha$ line}

The small FWHM measured for the Iron-K$\alpha$ line in our spectrum indicates that the line, as previously suggested by, e.g., \cite{Oosterbroek1996} and \cite{King2015}, is likely produced far away from the central BH.

While a Gaussian line still provides the best fit to our data around 6.4 keV, both the \textsc{diskline} and \textsc{Laor} models provide an inner disk radius (where the line is produced) of $R_\textrm{in} \sim 300 \,R_\textrm{g}$ (the value is unconstrained with \textsc{diskline}). 
\textsc{relxill} returns an inner radius of the emitting region of $R_\textrm{in} \sim 200\,R_\textrm{g}$, leaving structured residuals in the iron line region. These fits indicate that the iron line is not produced near the BH. These residuals are probably related to an edge that is likely due to absorption rather than to reflection (but see \citealt{Walton2016}).

\section{Spectral modelling - Model 2: Reprocessed Compton spectrum: the \textsc{MYTORUS} model}\label{sec:model2}

From the results obtained in the previous sections, it appears that: (i) the reprocessed/reflected emission dominates the X-ray broad band spectrum of V404 Cyg; (ii) heavy absorption affects the spectral shape of the source; (iii) partial covering of the central X-ray source is necessary to obtain good fits to the data. 
We note that both the absorption model \textsc{tbnew} and the reflection model we adopted\footnote{Calculated within \textsc{COMPPS}, which calculates the reflection spectrum according to \citealt{Magdziarz1995}}, (as basically all other reflection and absorption models available in \textit{XSPEC}), do not take into account the scattering associated with both absorption and reflection that becomes relevant already for column densities above a few $\times$10$^{23}$ atoms/cm$^2$ and thus significant for columns like those measured here, $\gtrsim \times$10$^{24}$ atoms/cm$^2$ (e.g. \citealt{Rybicki1979}). The effects of scattering in $\sim$ Compton-thick material affects the entire energy spectrum due to the  Klein-Nishina effect and should be carefully modelled in order to obtain reliable luminosities.

\textbf{\subsection{Model set-up}}

The observational facts listed above suggest that the properties of V404 Cyg closely resemble those of obscured AGN. Hence, we fitted to our data the \textsc{MYTORUS} model, \citep{Murphy2009, Yaqoob2012}, a spectral model describing a toroidal reprocessor that is valid from the Compton-thin to the Compton-thick regime. 
Even though MYTORUS was designed specifically for modelling the AGN  X-ray spectra, its use is not restricted to any system size scale and therefore can be applied to any axis-symmetric distribution of matter centrally-illuminated by X-rays. 

An extensive description of the basic properties of \textsc{MYTORUS} and its components is given in Appendix \ref{App:AppendixA}. 
The model expression that we used in this work is the following in \textsc{xspec}: 
\bigskip

\begin{quote}
\textsc{Model =  constant1 * tbnew1 * (constant2 * compTT1 + compTT2 * mytorus\_Ezero + constant3 * mytorus\_scattered1 + constant4 * mytorus\_scattered2 + (gsmooth * (constant5 * mytl1 + constant6 * mytl2)) + zgauss) + tbnew2 * powerlaw}
\end{quote}
\noindent
This expression depends on a relatively small number of free parameters (listed in Tab. \ref{tab:parameters2}): the interstellar column density \textsc{tbnew1} and \textsc{tbnew2}, the optical depth $\tau$, the constant factors weighting the contribution of the different components of the model (namely: \textsc{constant2}, \textsc{constant3} and \textsc{constant4}), the average column density N$_{H}$Z, and the line-of-sight column density N$_{H}$S, the centroid energy, FWHM and normalization parameters related to the zgauss component, and the power law parameters. \textsc{constant1} accounts for the relative normalizations of the spectra from the different instruments, and it is equal to 1 for \textit{Swift}/XRT and reported in Tab. \ref{tab:parameters2} for INTEGRAL/JEM-X and INTEGRAL/ISGRI.

In our fits, the optical depth $\tau$ is tied across all the components to the same (variable) value. 
The different components of the source spectrum are allowed to vary thanks to the constant factor preceding each of them (\textsc{constant2}, \textsc{constant3} and \textsc{constant4}). The constant associated to the fluorescent line spectra (\textsc{constant5} and \textsc{constant6}) are tied to the correspondent constants of the scattered spectra (\textsc{constant3} and \textsc{constant4}), as the line flux must be consistent with the scattered flux. 
The line-of-sight column density N$_{H}$S is tied across all the scattered components (continuum and line spectra), and can be either tied to or independent from the average column density N$_{H}$Z (related to the transmitted spectrum, i.e. the zeroth-order continuum, see App. \ref{App:AppendixA}).

\textsc{MYTORUS} must be used setting the same abundances and cross section used to produce all the \textsc{MYTORUS} model tables. Therefore, we used the cross section by \cite{Verner1996} and the abundances by \cite{Anders1989} instead of those by \cite{Wilms2000} that we used in Sec. \ref{sec:model1}. This implies a change in the ISM column density that we need to use with \textsc{MYTORUS}. Since the fit seems to be stable against small fluctuations of the ISM column density, we left this parameter free to vary, making sure that it did not drift to values inconsistent with those reported by \cite{Kalberla2005}.

\bigskip

\subsection{Fitting strategy}

Following \cite{Yaqoob2012}, we initially fitted our data set only above 10 keV. This allows to establish if the high-energy emission is dominated by the scattered emission or by the transmitted spectrum through the reprocessor (i.e., the zeroth-order continuum), which can never be zero. The best fit thus obtained shows that the transmitted spectrum dominates the emission above 10 keV. Since the-zeroth-order continuum depends on the electron temperature T$_{e}$ of the illuminating continuum, on the optical depth $\tau$ and on the average column density N$_{H}$Z, this best fit provides initial constraints on these parameters: $\tau$ = 0.9$\pm$0.2, T$_{e}$ = 27$_{-2}^{+3}$keV and N$_{H}$Z = (2.4$_{-2}^{+3}$)$\times$10$^{24}$cm$^{-2}$\footnote{Note that we are initially fitting only the high-energies (above 10keV) and as a consequence the column density is only poorly constrained in this exploratory fit.}. We kept the seed photon temperature fixed at 0.1 keV during the fit, however it stays consistent with 0.1 keV (while drifting to even lower values), even when left free to vary.
Since the electron temperature must be fixed when fitting \textsc{MYTORUS} to a particular data set, we fixed T$_{e}$ to 28~keV for all the following steps. This implies using the correct Monte-Carlo table, produced for a Compton illuminating spectrum with electron temperature 28 keV (see \citealt{Yaqoob2012}).

Being a rather complex model, \textsc{MYTORUS} can cause a spectral fitting degeneracy: two completely different model  configurations (e.g. either scattered emission or transmitted continuum dominating the spectrum) could describe equally well the same energy spectrum. However, the fact that the zeroth-order continuum dominates the high-energy emission above 10 keV in our case provides useful constraints to select the most appropriate model. In particular, any configuration where the scattered emission dominates over the transmitted one above 10 keV is to be discarded. 
Furthermore, when the zeroth-order continuum dominates the high-energy emission above 10 keV, the Compton-scattered continuum and Iron-K$\alpha$ line emission must be dominated by photons originating from back-illumination of the reprocessor and then reaching the observer along paths that do not intercept the Compton-thick structure. In other words, the structure must be clumpy, allowing a reflection continuum to reach the observer either from the far inner side of a toroidal structure or from an extended and dispersed distribution of matter. In particular, with the zeroth-order continuum dominating the emission above 10 keV, the radiation from \textit{back-illumination} of the reprocessor will dominate over the emission from reflection on the far inner side of the scattering torus \citep{Yaqoob2012}.

This configuration  can be modelled through MYTORUS used in the decoupled configuration (corresponding to the expression given above), in which the Compton-scattered continuum is composed of a face-on and an edge-on component (\textsc{mytorus\_scattered1} and \textsc{mytl1}, and \textsc{mytorus\_scattered2} and \textsc{mytl2}, respectively), each of which can be varied independently of the zeroth-order continuum. This setup can mimic a clumpy, patchy structure, axis-symmetric but not necessarily with a toroidal geometry. The inclination angle parameters in the \textsc{mytorus\_scattered1} and \textsc{mytorus\_scattered2} components of the decoupled model are fixed at 0$^{\circ}$  and 90$^{\circ}$, respectively, and are not related to the actual orbital inclination of the system. In this configuration, the inclination angle  determines only if the scattered emission intercepts ($\theta$ = 90$^{\circ}$) or not ($\theta$ = 0$^{\circ}$) the reprocessor before reaching the observer. Consequently, for the reasons given above, \textsc{mytorus\_scattered2} and \textsc{mytl2} components must dominate over the \textsc{mytorus\_scattered1} and \textsc{mytl1} ones. 
We decided to leave the average column density N$_{H}$Z, and the line-of-sight column density N$_{H}$S independent from each other, as in the presence of non-uniform and high column density absorbing material local to the source one should expect differences in the column density intercepted by the line of sight and the overall column density. 

\bigskip

\begin{figure*}
\centering
\includegraphics[width=1\textwidth]{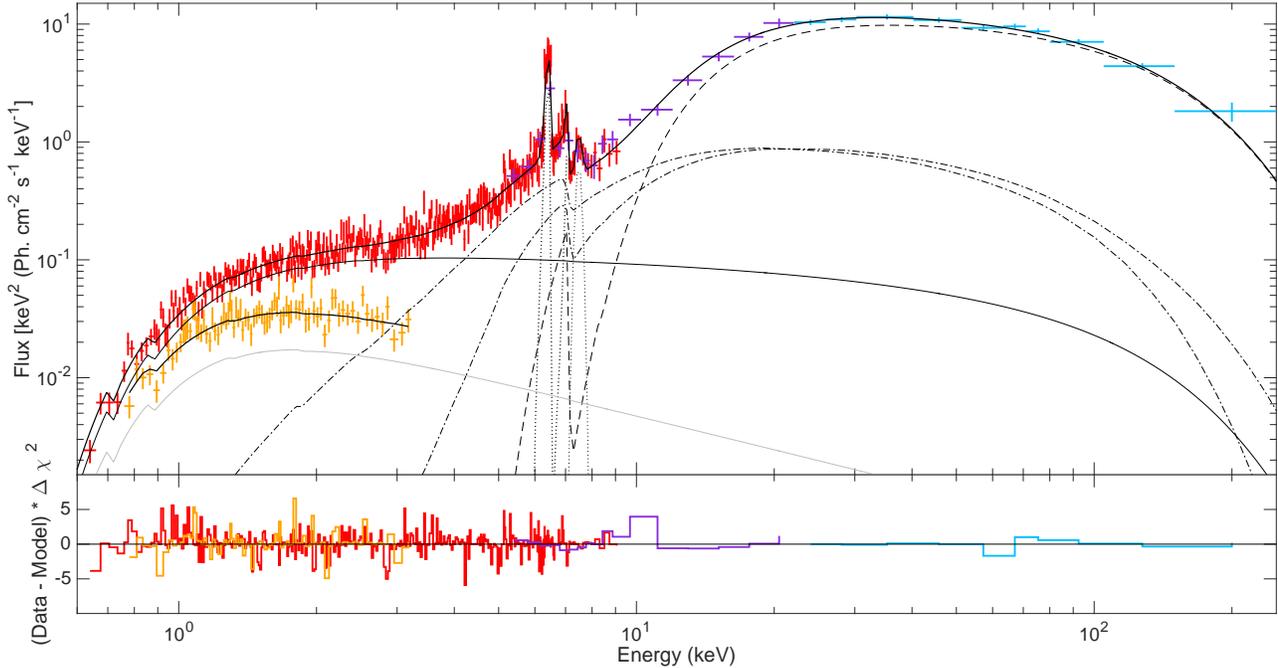}
\caption{\swift/XRT + \inte/JEM-X + \inte/ISGRI spectrum of V404 Cyg fitted with Model 2, i.e.  \textsc{MYTORUS} model in the decoupled configuration (see text for the details). \textit{Upper panel}: the thick black line marks the best fit to the data. The thin black line indicate the illuminating Compton continuum (\textsc{COMPTT1}), the dashed black line marks the zeroth-order continuum (\textsc{mytorus\_Ezero}), the dot-dashed lines mark the scattered continuum (\textsc{mytorus\_scattered1} and \textsc{mytorus\_scattered2}), the dotted line marks the fluorescent Fe line spectra (\textsc{mytl1} and \textsc{mytl2}) and the Gaussian line at $\sim$7.5 keV (\textsc{zgauss}). Finally, the solid grey line indicate the dust scattering halo emission. The color coding used to represent data points  is the same of Fig. \ref{fig:spettro}. 
A colour version of this figure is available on-line.}
\label{fig:spettro2}
\end{figure*}

We fitted the \textsc{MYTORUS} model in the decoupled configuration as described above to our full-band spectrum, following the same procedure we used previously, i.e. fitting the source data together with the dust scattering halo spectrum, in order to better constrain the halo spectrum slope (see Sec. \ref{sec:model1}). The best fit is shown in Fig. \ref{fig:spettro2}, the best fit parameters are given in Tab. \ref{tab:parameters2}, and the fluxes from each spectral component as well as the total source intrinsic flux are reported in Tab. \ref{tab:MYTORUS_fluxes}. The main difference between the results of this spectral modelling with respect to that described in Sec. \ref{sec:model1} (Model 2, case 3) is in the huge difference between the observed and the intrinsic source luminosity, which approaches a factor of 40. 

From the spectral modelling point of view, the main difference between Model 1 (see Sec. \ref{sec:model1}) and Model 2 is that (weak) residuals to the XRT data require the addition of a line at $\sim$7.5 keV, consistent with the Ni K-$\alpha$ line, expected in both AGN and binaries especially in the Compton-thick regime (\citealt{Yaqoob2011}). 
From this best fit we find that, similarly to Model 1, case 3 (see \ref{sec:Comp_refl}) the halo contributes only a little (0.3\%) to the total flux (source + halo) in the source region.
The resulting best fit is statistically favoured with respect to all the models described in the previous sections, with a $\chi^2$ = 469.58, 464 d.o.f.\ and null hypothesis probability equal to 0.419.

\begin{table}
\centering
\caption{Best fitting parameters from Model 2 (\textsc{MYTORUS}) described in Sec. \ref{sec:model2}. The  parameters marked with a * were initially allowed to vary, then they were fixed to their best-fit values for the sake of stability of the spectral fit. These parameters are then allowed to be free, one parameter at a time, in order to derive statistical errors. This is a procedure sometimes required when fitting MYTORUS to a data set, aimed at keeping control on the spectral parameter of a model that is rather complex compared to the majority of the models included in \textsc{XSPEC} (\citealt{Yaqoob2012}). The line fluxes are measured in the 0.6-10 keV energy band. The lines flux has been expressed in term of erg cm$^{-2}$s$^{-1}$ instead of equivalent width since it is not possible to unambiguously measure the equivalent width of the Fe K lines given the complexity of the continuum. In other words, it is difficult to determine what continuum should be referred  to the Fe K lines, therefore we decided to report the observed line flux. }\label{tab:parameters2}
\begin{tabular}{c c }

\hline
Parameter												& 		Value					 \\
\hline
\hline
constant2												&	(6$\pm$2)$\times$10$^{-4}$			\\
$\tau$													&	0.89$\pm$0.04								\\
T$_{e}$	[keV]											&	28 keV (fixed)								\\
K$_{COMPTT}$											&	34$_{-10}^{+13}$							\\
N$_{H}$Z [cm$^{-1}$]									&	(3.2$\pm$0.3)$\times$10$^{24}$ 	\\
constant3 (=constant5)									&	0.4$\pm$0.1						\\
N$_{H}$S [cm$^{-1}$]									&	0.8$\pm$0.1 $\times$10$^{24}$ 		\\
constant4 (=constant6)									&	0.5$\pm$0.2								\\

$\Gamma_\textrm{halo}^*$								&	2.96$\pm$0.04								\\
$K_\textrm{halo}^*$ (in the source region)				&	4.4$\pm$0.3$\times$10$^{-2}$				\\
cross-normalization constant (JEM-X)					&	1.30$\pm$0.08								\\
cross-normalization constant (ISGRI)					&	1.5$_{-0.1}^{+0.2}$							\\	
Fe K-$\alpha$ and K-$\beta$ Lines Flux [erg cm$^{-2}$s$^{-1}$]	&	(2.8$\pm$0.7)$\times$10$^{-10}$										\\
Line energy (Ni K-$\alpha$)$^*$	[keV]					&	7.52$\pm$0.09 							\\
Ni K-$\alpha$ Line Flux$^*$	[erg cm$^{-2}$s$^{-1}$]		&	(3.7$\pm$0.9)$\times$10$^{-11}$									\\
\hline
$\chi^2$												& 	469.58										\\	
Degrees of freedom										&   464											\\
Null hypothesis probability 							&  	4.19$\times$10$^{-1}$											\\

\hline
\end{tabular}
\end{table}

\begin{table}
\centering
\caption{Fluxes obtained in the 0.6-200 keV energy band from Model 2 (see Sec. \ref{sec:model2}). Save for the source flux labelled as \textit{intrinsic} (which correspond to the intrinsic flux emitted before being reprocessed by the local absorber as well as by the interstellar medium), all the reported fluxes are not corrected for the absorption local to the source, therefore they must be interpreted as observed fluxes corrected for the inter stellar medium absorption only. The \textit{Compton continuum} flux is the flux fraction intercepted by the line of sight without being reprocessed by the local absorber. The \textit{scattered continuum} fluxes include the contribution of the Iron lines. For reference, the Flux corresponding to the Eddington luminosity equals 1.6$\times 10^{-6}\, \ergcms$  for a stellar mass black hole of 9 M$\odot$ and at 2.39 kpc like V404 Cyg.}\label{tab:MYTORUS_fluxes}
\begin{tabular}{c c }		
\hline	  
Component 							& 	Flux [erg/cm$^{2}$/s]	\\
\hline
\hline
Source 								& 3.65 $\times 10^{-8}$		\\
Source (intrinsic)			    	& 1.31 $\times 10^{-6}$  \\
Compton continuum 					& 7.61 $\times 10^{-10}$ 	\\
zeroth-order continuum 				& 2.98 $\times 10^{-8}$ 	\\
Scattered continuum including Fe lines (0$^{\circ}$)  	& 3.09 $\times 10^{-9}$ 	\\
Scattered continuum including Fe lines (90$^{\circ}$) 	& 2.78 $\times 10^{-9}$  	\\
Halo (in the source region)			& 1.17 $\times 10^{-10}$ 	\\

\hline
\end{tabular}
\end{table}

\bigskip

\section{Discussion}

We analysed a simultaneous \inte\ and \swift\ spectrum of the black hole binary V404 Cyg obtained during the 2015 Summer outburst, when the source was in a plateau, reflection dominated state. This is the first time that an X-ray spectrum of V404 Cyg is available simultaneously over such a wide energy range (0.6--250\,keV).

The broad-band X-ray energy spectrum of V404 Cyg is remarkably similar to the typical spectra of obscured AGN, where the primary emission is absorbed and reprocessed by high column density of gas. The spectrum analysed in this work can be well described by a combination of direct Compton emission produced by hot electrons ($\sim$65\,keV, see Tab. \ref{tab:parameters}) in a optically translucent material ($\tau \sim$ 1.2) up-scattering low-temperature photons, and reflected emission including a narrow line centred at 6.4 keV. The presence of a high column density neutral absorber (equivalent $N_\textrm{H} \approx 1.4 \times 10^{24}\, \textrm{cm}^{-2}$) covering about 85 per cent of the central source is necessary to describe the soft X-ray emission (below 10 keV). The prominent reflection hump especially evident in the \inte\ IBIS/ISGRI and JEM-X spectra, is indicative of Compton back-scattering of photons by Compton-thick material around the source. 
The very low FWHM of the Iron-K$\alpha$ line suggests that the reflection takes place far away from the central black hole, as previously observed during the 1989 X-ray outburst of the source (\citealt{Oosterbroek1997}). 
The heavy absorption we derive ($N_\textrm{H} \sim$ $1.4\times$10$^{24}\,\textrm{cm}^{-2}$) is consistent with the values measured during the 1989 outburst.

A more sophisticated modelling of the broad-band X-ray spectrum that takes into account both the complex geometry of the absorber and the presence of heavily reprocessed emission, shows that the best description of the spectrum is given by a bright point source whose emission is scattered and reflected by a patchy toroidal reprocessor surrounding it. 
In this case the illuminating spectrum can still be well-described by a Compton spectrum with optical depth $\tau \sim 0.9$, smaller than in the case where scattering is not taken into account. The electron temperature is also smaller with respect to the previous case, but consistent with the average electron temperature measured from INTEGRAL/ISBIS/ISGRI data (\citealt{Sanchez-Fernandez2016}). 
This difference can be ascribed to the fact that, in the presence of heavy absorption, the effect of scattering significantly affects the source high-energy curvature by producing a strong transmitted component. The result is a blue-shift of the high-energy spectral roll-over, that can be erroneously ascribed to a higher electron temperature when the effects of scattering are  not properly modelled.
The average column density of the toroidal reprocessor is nearly in the Compton-thick regime ($N_\textrm{H} \approx 3 \times 10^{24}\, \textrm{cm}^{-2}$), while the column density in the direction of the line of sight is about a factor of four smaller ($N_\textrm{H} \approx 0.8 \times 10^{24}\, \textrm{cm}^{-2}$). This is expected in a scenario where a local, non uniform (patchy) and highly variable reprocessor heavily affects the emission from the central source, as the Comptonized emission, the transmitted emission and the scattered emission (which includes what is typically referred to as reflected emission) do not necessarily experience the same absorption. In particular, while the transmitted emission (i.e., the zeroth order continuum, which has essentially lost any direction information being the result of multiple scattering events) carries the effects of the overall average column density, the scattered emission is mostly affected by the line of sight column density,  which can be significantly larger then the overall one. This can happen, for instance, in the situation where a Compton-thick clump of cold material is intercepting the line of sight.
As expected in the presence of heavy absorption, the high-energy spectrum (i.e. above 10 keV) is dominated by the reprocessed emission. A large fraction of such emission comes from photons scattered multiple times in the local absorber, while the remaining fraction is due to photons scattered through back-illuminated matter in the reprocessor, and to photons reflected off the far inner side of the reprocessor and then reaching the observer without being further scattered. 

The line emission associated to the Iron-K$\alpha$ and Iron-K$\beta$ transitions is not resolved in \textit{Swift}/XRT and hence are seen as one single line. The Ni-K$\alpha$ line is also unresolved by XRT. The relative observed flux of the Iron lines and of the Nickel line is in agreement with predictions by \cite{Yaqoob2011}, with the Nickel line flux being a factor $>7$ fainter than the Iron lines flux. The fact that the Nickel line is not significantly detected in the \textit{Swift}/XRT spectrum when fitting it using Model 1 (case 3) can be a consequence of a cruder modelling of the region around 7 keV, where the iron edges affect the most of the spectrum.

The most relevant difference that we found between the two models we considered is the inferred intrinsic luminosity of the central X-ray source. On the one hand, a simple reflected Comptonized spectrum (Model 1, case 3, see Sec. \ref{sec:Comp_refl}), which does not consider scattering, but only pure reflection (i.e. the reflecting material is assumed to be characterized by infinite optical depth), returns a flux of 4.46 $\times 10^{-8}$ erg/cm$^{2}$/s, corresponding to $\sim$3\,\% the Eddington luminosity for a 9M$_{\odot}$ black hole. 
On the other hand, a more complex modelling of the reprocessed emission that takes into account the scattering processes (Model 2, see Sec. \ref{sec:model2}) gives an intrinsic flux of 1.31 $\times 10^{-6}$ erg/cm$^{2}$/s, corresponding to $\sim$ the Eddington luminosity for V404 Cyg.
As mentioned earlier, the data considered in this work correspond to a \textit{plateau} phase in the light curve of V404 Cyg (see \citealt{Sanchez-Fernandez2016}) that occurred in between two of the major flares seen in the 2015 Summer outburst (flare peak observed on MJD 57194.11 and MJD 57194.31, respectively). We estimated the ratio R$_{fl/pl}$ of the average flare-peak flux to the average plateau flux from the INTEGRAL/ISGRI light curve in the 20-200 keV energy band \citep{Kuulkers2015c}  obtaining  R$_{fl/pl} = 10\pm 2$. Then we compared such ratio with the ratio R$_{Int/Obs}$ of the intrinsic to the observed flux in the 20-200 keV energy band using the best fit to the data based on MYTORUS, which gave R$_{Int/Obs} = 8.6 \pm 0.9$. The fact that the two ratios are consistent indicate that the plateau where our data set comes from could be the result of a short-lived almost complete obscuration of the central source in a very bright phase, rather then an actual decrease in the emitted flux. This is in agreement with the results of the spectral modelling,  which required heavy absorption and, consequently, a significant amount of reprocessed emission, regardless of the details of the model used.

\bigskip

\noindent \citet{Zycki1999} and \citet{Oosterbroek1996} reported on the high variability of an 
intrinsic absorption component  based both on the spectral analysis of the source and on the time scales over which the absorption was changing, which suggests fast movement and/or fast changes in the physical properties (e.g. optical depth) of material within the system. 
\citet{Zycki1999} suggested that the presence of heavy (and variable) intrinsic absorption is likely the main reason why the energy spectrum of V404 Cyg almost never resembles the spectrum typical of any of the spectral-timing canonical states of BHBs in outburst.

Our results are in good agreement with those obtained with \textit{Ginga}. 
The \swift/XRT sensitivity to the low energies, together with the \inte\ broad-band coverage, allowed to study in detail the combined effects of  heavy absorption and strong reflection in the X-ray spectrum of V404 Cyg.
The X-ray central engine is probably hidden beneath a layer of complex, heavily absorbing material that substantially suppresses the source intrinsic spectrum, making it hard to recover it.
The in-homogeneity of the absorber is such that occasionally, the observer can get a glimpse of the un-obscured source, together with the reflected spectrum. A similar scenario was proposed to explain the properties of V4146 Sgr during the 1999 outburst \citep{Revnivtsev2002}. After a Super-Eddington phase, the system ejected a significant amount of matter that was responsible for heavy and non-homogeneous absorption and intense reflection. The V4641 Sgr X-ray spectrum after the outburst peak was remarkably similar to that of a Type-2 AGN \citep{Morningstar2014}. 

The derived electron temperature of the Comptonizing medium (for both Model 1 and Model 2) is consistent with the  
results from other authors for V404 Cyg during the 2015 outburst (\citealt{Natalucci2015} and \citealt{Rodriguez2015}, \citealt{Sanchez-Fernandez2016}) and with what has been observed in other, more canonical
transient BHBs (e.g., Cyg X-1, \citealt{Sunyaev1979}, GX 339-4,  \citealt{DelSanto2008}, \citealt{Motta2009}, and GRO J1655-40, \citealt{Joinet2008}).
Low temperature seed photons required to produce the direct Comptonized spectrum are normally found in black hole binaries at low luminosities (see, e.g., \citealt{Dunn2011}). 
Such seed photons would come, in our case, from either a (cool) heavily absorbed accretion disk truncated at large radii (e.g., \citealt{Done2007}) and/or from synchrotron self-Compton emission by non-thermal electrons in the hot Comptonizing medium (see, e.g., \citealt{Poutanen2014}, \citealt{Kajava2016}).

Since we do not find evidences of a soft component, such as a disk black-body,
our data alone do not allow to unambiguously determine the origin of the Compton seed photons. \citet{Zycki1999} reported the detection of a short-lived disk dominated state during the 1989 outburst of V404 Cyg. However, given the extreme luminosities reached during the 1989 outburst -- comparable to those observed in 2015 --  it is reasonable to assume that a dust scattering halo (\citealt{Beardmore2015}, \citealt{Vasilopoulos2015} and \citealt{Heinz2016}) formed also back in 1989. 
Our results show that the presence of this halo does not contaminate the overall emission of the source to a significant level in the \swift/XRT observation. 
However, given the large field of view (1.1$\times$2.0 $^{\circ 2}$ FWHM)  of the \textit{Ginga} collimated proportional counter array (LAC, \citealt{Turner1989}), \textit{Ginga} would have not been able to disentangle the source emission from the halo emission. 
Therefore, it is possible that the soft emission ascribed to an accretion disk in \textit{Ginga} data by \citet{Zycki1999} is in reality soft emission from the halo.

\subsection{V404 Cyg: an obscured super-Eddington AGN-analogue}

Both significant reflected emission (see, e.g., NGC 7582, \citealt{Bianchi2009} and \citealt{Rivers2015}) and the effects of a patchy, neutral absorber (see, e.g., the case of NGC 4151, \citealt{Zdziarski2002a}, \citealt{Derosa2007} and NGC 1365, \citealt{Risaliti2005}) are sometimes seen in obscured AGN, where the variability of the absorber is thought to be responsible of most of the variability from the source. 

High values of reflection fractions in AGN are normally ascribed to the fact that the source of the illuminating continuum is no longer visible/active, i.e. because of intervening partially covering absorption or because the source switched off, and the only radiation seen is the reflected one (see, e.g., \citealt{Rivers2015}, but see, e.g., \citealt{Miniutti2004} for a different scenario).
In both cases, the reflection amplitude is bound to increase significantly. V404 Cyg also showed substantial reflected emission, however in this system the scenario is probably slightly different from that of an obscured and/or reflected AGN, since the illuminating continuum can be directly observed, though largely absorbed, together with the (dominating) reflected emission. This suggests that the spectrum is
a combination of Comptonized continuum and reprocessed emission, likely produced in different areas of the system (i.e., close to the central black hole and further out, respectively).

According to the unified model of AGN (\citealt{Antonucci1993}, \citealt{Urry1995}), the central BH is always surrounded by an axis-symmetric parsec-scale torus. Furthermore, a large fraction of AGN show clear evidence of absorption in the soft X-ray band, interpreted as material, either neutral or ionized, on the line of sight (\citealt{Turner2009}). Recent finding have indicated that this absorber is most likely non homogeneous and located close to the central black hole (e.g., \citealt{Risaliti2005}), at a smaller distance than the dust torus. In addition, it has been found that in several AGN, the reflection components in the X-ray spectra are significantly stronger than expected for reflection off gas with the same column density measured from the absorption features (e.g., \citealt{Guainazzi2005}). This is only possible if a thick reflector close to the BH and well-within the parsec scale torus, with column densities exceeding $N_\textrm{H} = 10^{24}\,\textrm{cm}^{-2}$,  covers a large fraction of the solid angle around the source (\citealt{Ghisellini1994}).
In the spectrum of V404 Cyg we detected, together with a (weak) direct Comptonized spectrum, both high reflection and the signatures of heavy, non-homogeneous absorption, all effects pointing to the presence of non-uniform shielding material local to the source on the line of sight. 

The similarities between the properties of V404 Cyg and those of some AGN, suggest that the accretion configuration in the former might be very close to that expected in obscured but intrinsically luminous AGN, accreting at high accretion rates, where the inner accretion flow is well described by the \textit{slim disk} model (\citealt{Abramowicz1988}) and the central engine is thought to be partially or completely obscured by an absorber (the flared inner accretion disk) located close to the central black hole and internal w.r.t. the dust torus; in this case both reflection and absorption play a key role in shaping the broadband energy spectrum. 

Simulations show that both in stellar mass accreting BH and in AGN, high (super-Eddington) accretion rates can develop strong radiation forces able to sustain a thick accretion flow that might form at times a non-homogeneous (i.e., clumpy) mass outflow, launched within a few hundreds of $\mathrm{R}_\textrm{g}$ from the black hole (see, e.g., \citealt{Takeuchi2013}).
In addition, similar scenario has been suggested also for the ultra-luminous X-ray sources (e.g., \citealt{PLF07}).
In other words, the geometrically thin, optically thick accretion disk -- the launching site of the winds seen in the optical band at thousands of $\mathrm{R}_\textrm{g}$ from the BH (\citealt{Munoz-Darias2016}) -- puffs-up in its inner tens to hundreds of $\mathrm{R}_\textrm{g}$, becoming a geometrically thick accretion flow, sustained by the radiative forces that develop as a consequence of the high-accretion rates. This thick accretion flow then fragments out at a certain distance from the disk plane, forming high-density Compton-thick clumps of material, which could be responsible for the high intrinsic, non-homogeneous absorption seen in V404 Cyg, as well as for the intense reflected emission.
When the inclination is high enough -- like in the case of V404 Cyg -- this inner \textit{slim disk} is able to shield the innermost region of the accretion flow, preventing the radiation to directly reach the observer most of the time. The observed emission from these objects is therefore expected to be dominated by scattered/reflected radiation. Such a high-accretion rate regime is rarely observed in BHBs, but it is inferred to be present in about 1 per cent of high redshift optically selected AGN (\citealt{Luo2015}). 
The first example of high quality X-ray spectrum of a Super-Eddington AGN has been presented in \citet{Lanzuisi2016}, where one of the plausible scenarios that can explain the data is an intrinsic emission strongly reprocessed through absorption and reflection in partially covering Compton-thick material. In this case the obscuration of the AGN is not due to a distant parsec-scale torus, but rather to the inner accretion flow itself that under the strong radiation pressure puffs up into the slim disk configuration.

The alternating phases of high and low luminosities observed during the 2015 outburst of V404 Cyg (\citealt{Natalucci2015}, \citealt{Rodriguez2015}) suggest that \textit{V404 Cyg might have been accreting erratically or even continuously at super-Eddington rates}, while being partly or completely obscured by a in-homogeneous, high-density layer of neutral material local to the source (similarly to what happened to V4146 Sgr, \citealt{Revnivtsev2002}).
In this context, the fact that the reprocessed emission almost dominates the entire spectrum, implies that the emitted luminosity can be order of magnitudes higher than what is directly measured (see, e.g., \citealt{Murphy2009}), as our results suggest. 
This has strong implications in the context of X-ray/radio correlations (e.g., \citealt{Gallo2012}). 
The large difference between observed and measured X-ray flux should be taken into account carefully, since while the X-ray emitting region could be almost completely obscured, the radio emitting region is most likely always  visible as it is probably emitted from a few to tens of  $\mathrm{R}_\textrm{g}$ away from the accretion disk mid-plane.

\section{Summary and conclusions}								

We have analysed unique simultaneous \inte\ and \swift\ observations of the black hole candidate V404 Cyg (GS 2023+338) during the 2015 summer outburst. We observed the source in a rare, long, plateau, reflection-dominated state, where the energy spectrum was stable enough to allow time-averaged spectral analysis. 

Fits to the source X-ray spectrum in the 0.6--200 keV energy range
revealed heavily absorbed, Comptonized emission as well as significant reprocessed emission, dominating at high energies (above $\sim$10\, keV).
The measured average high column density ($N_\textrm{H} \approx 1-3 \times 10^{24}\,\textrm{cm}^{-2}$) is likely due to absorption by matter expelled from the central part of the system.  The overall X-ray spectrum
is consistent with the X-ray emission produced by a thick accretion flow, or \textit{slim disk}, similar to that expected in obscured AGN accreting at high accretion rates (i.e. close to the Eddington rate), where the emission from the very centre of the system is shielded by a geometrically thick accretion flow. 

We therefore suggest that in some of the low-flux/plateau states detected between large X-ray flares during the 2015 outburst, the spectrum of V404 Cyg is similar to the spectrum of an obscured AGN. 
Given the analogy and the extreme absorption measured, we argue that occasionally the observed X-ray flux might be very different from the system intrinsic flux, which is almost completely reprocessed before reaching the observer.
This may be particularly important when comparing the X-ray and radio fluxes, since the latter is likely always emitted sufficiently far away from the disk mid-plane and therefore never obscured.

Given the fact that accretion should work on the same principles in BHBs and AGN, once a suitable scale in mass is applied, detailed studies of V404 Cyg and stellar mass black holes with similar characteristics could help in shading light on some of the inflow/outflow dynamics at play in some, still poorly understood, classes of obscured AGN.

\bigskip

SEM acknowledges the anonymous referee whose useful comments largely contributed to improve this work. 
SEM acknowledges the University of Oxford and the Violette and Samuel Glasstone Research Fellowship program and ESA for hospitality. SEM also  acknowledges Rob Fender, Andy Beardmore and Robert Antonucci for useful discussion. JJEK acknowledges support from the Academy of Finland grant 268740 and the ESA research fellowship programme. SEM and JJEK acknowledge support from the Faculty of the European Space Astronomy Centre (ESAC). EK acknowledges the University of Oxford for hospitality. MG acknowledges SRON, which is supported financially by NWO, the Netherlands Organisation for Scientific Research. 
This work is based  on  observations  with  INTEGRAL,  an  ESA  project  with  instruments  and  science  data centre funded by ESA member states (especially the PI countries: Denmark, France, Germany, Italy, Switzerland, Spain), and with the participation of Russia and the USA.



\bibliographystyle{mn2e.bst}
\bibliography{biblio} 


\appendix

\section{The MYTORUS model - a brief \textit{vademecum}} \label{App:AppendixA}

The MYTORUS model was designed specifically for modelling the X-ray spectra of active galaxies, however, it is not restricted to any absolute size scale. It adopts a tube-like, azimuthally-symmetric torus, where \textit{c} is the distance from the centre of the torus to the centre of the “tube”, and \textit{a} is the radius of the tube. In \textsc{MYTORUS} only the ratio, \textit{c/a}, is relevant, which makes the model essentially scale-free. Hence, \textsc{MYTORUS} can be applied to any toroidal distribution of matter that is centrally-illuminated by X-rays.

The \textsc{MYTORUS} model consists in a number of components, each devoted to describing a particular part of the radiation reprocessing:

\begin{itemize}
\item The \textit{zeroth-order continuum}, which is formed by the collection of photons that leave the absorbing medium without interacting with it, i.e. the photons are neither scattered or absorbed. The zeroth-order continuum is a purely line-of-sight quantity, in the sense that it does not depend on the geometry or covering fraction of the material away from the line of sight. In other words, it is a fraction of the input spectrum at a given energy, which depends on a energy-dependent optical depth.

\item The  \textit{Scattered continuum}, which is the collection of all escaping photons that have been scattered in the medium at least once. The net scattered spectrum for a cold medium at a given energy then depends on the input spectrum at all higher energies. The fractional energy shift due to Compton scattering depends on the initial energy of the photon itself and is larger for higher energies. 

\item The \textit{Fluorescent emission line spectrum}, which is the collection of photon absorbed above the K-edge threshold energy of an atom or ion, that can trigger the removal of an electron and the consequent decay of an upper level electron, causing the emission of a  fluorescent line. The absorbed photons can be either photons absorbed before being scattered (and then leaving the absorbing medium without being scattered), or be photons that have been scattered before leaving the medium.  

\end{itemize}

These three components can be combined in order to obtain different geometries of the toroidal reprocessor. The results obtained in Sec. \ref{sec:model1} show that the presence of a patch/clumpy absorber local to the source is necessary to explain the broad-band X-ray spectrum. Therefore, following the indication of \cite{Yaqoob2012}, we fitted to our data the \textsc{MYTORUS} model in its \textit{decoupled} version, which allows to take into account the non uniformity of the reprocessor. 

The model expression that we used in this work is the following: 
\begin{quote}

\textsc{Model =  constant1 * tbnew1 * (constant2 * compTT1 + compTT2 * mytorus\_Ezero + constant3 * mytorus\_scattered1 + constant4 * mytorus\_scattered2 + (gsmooth * (constant5 * mytl1 + constant6 * mytl2)) + zgauss) + tbnew2 * powerlaw}

\end{quote}

\textsc{Constant1} is the instrumental normalization, aimed at taking into account differences in the normalization of the three instruments we use, and it is fixed at 1. 
\textsc{tbnew1} and \textsc{tbnew2} are neutral, uniform absorbers, with column density frozen to the interstellar values in the direction of V404 Cyg. \textsc{tbnew1} is applied to the source overall spectrum, while \textsc{tbnew2}, tied to \textsc{tbnew1}, is applied to the \textsc{powerlaw} component aimed at describing the halo emission.

\textsc{COMPTT1} and \textsc{COMPTT2} are both associated to the source illuminating spectrum, but while the former is intended to model the unobscured illuminating spectrum (the emission that reaches the observer through the 'holes' in the patchy absorber), the latter is attenuated by the effects of the toroidal reprocessor (\textsc{mytorus\_Ezero}), giving rise to the zeroth-order continuum. 
The main parameters of the COMPTT are the seed photons temperature $T_\textrm{0}$, the electron temperature $T_\textrm{e}$, the optical depth $\tau$ and the normalization. As we did in Sec. \ref{sec:model1}, we fixed $T_\textrm{0}$ to 0.1 keV. The main parameter of \textsc{mytorus\_Ezero} is the average column density  $N_\textrm{H}$Z. 
The reason why we used COMPTT and not COMPPS is that MYTORUS is currently designed to allow either a power law intrinsic spectrum, or a Compton spectrum as described by \textsc{COMPTT}. Even if small differences between \textsc{COMPTT} and \textsc{COMPPS} are to be expected, we can reasonably assume that our results will be not affected by them, since the reprocessed emission, which dominates the emission, is not too sensitive to the details of the illuminating spectrum. 

\textsc{mytorus\_scattered1} and \textsc{mytorus\_scattered2} describe two different parts of the scattered spectrum: \textsc{mytorus\_scattered1} accounts for the emission scattered on the far inner side of the toroidal reprocessor, that then reaches the observer without being further scattered; \textsc{mytorus\_scattered2} accounts for the emission from material back-illuminated from the central source (for details, see \citealt{Yaqoob2012} and in particular their fig. 2). The main parameters of \textsc{mytorus\_scattered1} and \textsc{mytorus\_scattered2} are the column density along the line of sight  $N_\textrm{H}$S, the inclination angle parameter $\theta$ (which denotes the inclination angle between the line of sights and the axis of the reprocessing torus, and determines if the line of sight intercepts or not the reprocessor) and the optical depth $\tau$. 

\textsc{mytl1} and \textsc{mytl2} are the fluorescent line spectra associated to \textsc{mytorus\_scattered1} and \textsc{mytorus\_scattered2}, respectively. The main parameters of these components are the column density along the line of sight $N_\textrm{H}$S and the optical depth $\tau$ and the inclination angle parameter, which in each fluorescent line spectrum component is tied to the correspondent scattered spectrum component. 
\textsc{gsmooth} is typically applied to the fluorescent line spectra and takes into account the possible velocity broadening of the lines, however our best fit did not require any significant broadening, therefore the broadening parameter was fixed to zero.

\textsc{zgauss} is an additional line  added to the final model in order to describe residuals found in the XRT source spectrum around the Ni K-$\alpha$ line ($\sim$7.5 keV).

\bsp	
\label{lastpage}
\end{document}